\documentclass[submission, Proceedings]{SciPost}

\hypersetup{
    colorlinks,
    linkcolor={red!50!black},
    citecolor={blue!50!black},
    urlcolor={blue!80!black}
}

\usepackage[bitstream-charter]{mathdesign}
\usepackage{todonotes}
\usepackage{caption}
\usepackage{subcaption}
\DeclareSymbolFont{usualmathcal}{OMS}{cmsy}{m}{n}
\DeclareSymbolFontAlphabet{\mathcal}{usualmathcal}

\begin{document}

\begin{flushright}{
LUXE-Proc-2021-001
}\end{flushright}

\begin{center}{\Large \textbf{
LUXE: A new experiment to study non-perturbative QED in $e^{-}$-laser and $\gamma$-laser collisions \\
}}\end{center}

% TODO: write the author list here. Use initials + surname format.
% Separate subsequent authors by a comma, omit comma at the end of the list.
% Mark the corresponding author with a superscript *.
\begin{center}
Ruth Jacobs\textsuperscript{1$\star$},
on behalf of the LUXE collaboration
\end{center}

\begin{center}
{\bf 1} Deutsches Elektronen-Synchrotron DESY, Notkestr. 85, 22607 Hamburg, Germany
\\
* ruth.magdalena.jacobs@desy.de
\end{center}

\begin{center}
\today
\end{center}

% For convenience during refereeing (optional),
% you can turn on line numbers by uncommenting the next line:
%\linenumbers

\definecolor{palegray}{gray}{0.95}
\begin{center}
\colorbox{palegray}{
  \begin{tabular}{rr}
  \begin{minipage}{0.1\textwidth}
    \includegraphics[width=22mm]{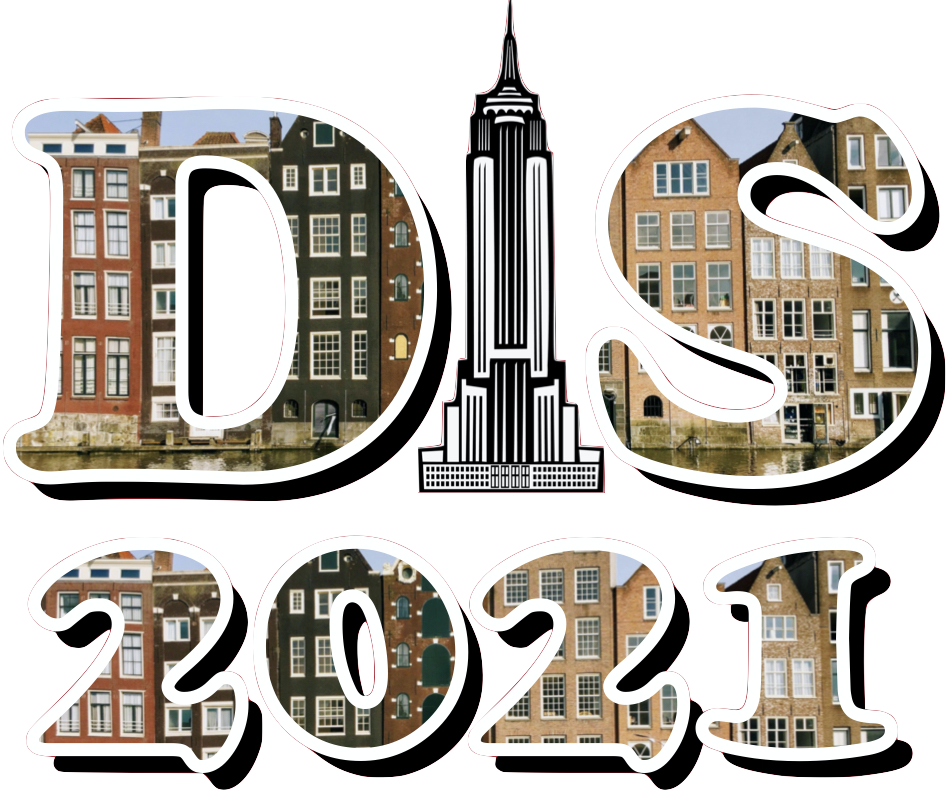}
  \end{minipage}
  &
  \begin{minipage}{0.75\textwidth}
    \begin{center}
    {\it Proceedings for the XXVIII International Workshop\\ on Deep-Inelastic Scattering and
Related Subjects,}\\
    {\it Stony Brook University, New York, USA, 12-16 April 2021} \\
    \doi{10.21468/SciPostPhysProc.?}\\
    \end{center}
  \end{minipage}
\end{tabular}
}
\end{center}

\section*{Abstract}
{\bf
% TODO: write your abstract here.
The LUXE experiment (Laser Und XFEL Experiment) is a new experiment in planning at DESY Hamburg using the electron beam of the European XFEL. LUXE is intended to study collisions between a high-intensity optical laser and up to 16.5 GeV electrons from the Eu.XFEL electron beam, or, alternatively, high-energy secondary photons. The physics objective of LUXE are processes of Quantum Electrodynamics (QED) at the strong-field frontier, where QED is non-perturbative. This manifests itself in the creation of physical electron-positron pairs from the QED vacuum. LUXE intends to measure the positron production rate in a new physics regime at an unprecedented laser intensity. Parasitically, the high-intensity Compton photon beam of LUXE can be used to search for physics beyond the Standard Model. }

%\vspace{10pt}
%\noindent\rule{\textwidth}{1pt}
%\tableofcontents\thispagestyle{fancy}
%\noindent\rule{\textwidth}{1pt}
%\vspace{10pt}

\section{Introduction}
\label{sec:intro}
% TODO: write your article here.
%The stage is yours. Write your article here.
%The bulk of the paper should be clearly divided into sections with short descriptive titles, including an introduction and a conclusion.

Quantum Electrodynamics (QED) is the most well-tested theory in physics. The success of QED is based on comparing ultra-precise perturbative calculations with experimental data. However there exists a regime in which perturbative calculations of QED break down, namely in the vicinity of a strong background field. If the field energy (i.e. the work by the field over one Compton wavelength) is larger than the rest mass of a virtual particle, the vacuum becomes polarized.  In strong-field QED this polarization is expected to manifest itself in the creation of physical electron-positron pairs from virtual electron-positron vacuum fluctuations \cite{Heisenberg:1935qt}. The critical field strength, $\mathcal{E}_{\text{crit}}$, required for this process is called the "Schwinger limit"~\cite{Schwinger:1951}. For an electrical field, the Schwinger critical field corresponds to $\mathcal{E}_{\text{crit}}=m_{\text{e}}^2c^3/e\hbar\approx 1.32 \times 10^{18}\;\text{V/m}$. In nature, strong-field QED plays a role, for example in the gravitational collapse of Black Holes~\cite{Ruffini:2009hg}, the propagation of cosmic rays~\cite{Nishikov}, or in processes on the surface of strongly magnetized neutron stars ~\cite{Kouveliotou:1998ze,Turolla:2015mwa,Kaspi:2017} \\ 

The LUXE experiment ~\cite{LUXECDR} proposed at DESY and the European XFEL (Eu.XFEL) in Hamburg and Schenefeld, Germany, is intended to study strong-field QED processes in collisions of a high-intensity optical laser and the 16.5 GeV electron beam of the Eu.XFEL, as well as collisions of the high-intensity optical laser and high-energy secondary photons. The strong background field is provided by the Terawatt laser-pulse and enhanced by the Lorentz boost of the electrons. This will allow LUXE to explore a previously uncharted intensity regime.\\

This article is structured as follows: Section \ref{sec:sfqed} gives an overview of strong-field QED processes and parameters. Section \ref{sec:luxesetup} introduces the proposed LUXE experimental setup, including the laser and particle detectors. Section \ref{sec:expresults} showcases the expected measurement results. Finally, the conclusion is presented in section \ref{sec:conclusion}.

\section{Strong-field QED processes at LUXE}
\label{sec:sfqed}
%Use sections to structure your article's presentation.

%Equations should be centered; multi-line equations should be aligned.
%\begin{equation}
%H = \sum_{j=1}^N \left[J (S^x_j S^x_{j+1} + S^y_j S^y_{j+1} + \Delta S^z_j S^z_{j+1}) - h S^z_j \right].
%\end{equation}

%In the list of references, cited papers \cite{1931_Bethe_ZP_71} should include authors, title, journal reference (journal name, volume number (in bold), start page) and most importantly a DOI link. For a preprint \cite{arXiv:1108.2700}, please include authors, title (please ensure proper capitalization) and arXiv link. If you use BiBTeX with our style file, the right format will be automatically implemented.

%All equations and references should be hyperlinked to ensure ease of navigation. This also holds for [sub]sections: readers should be able to easily jump to Section \ref{sec:another}.

%The experimental concept of LUXE is to collide a high-intensity optical laser pulse either directly with the 16.5 GeV electron beam of the Eu.XFEL linear accelerator, or with secondary high-energy photons, created by impinging the XFEL electron beam on a thin converter target, producing Bremsstrahlung photons in the GeV energy range. \\ 

Two parameters are commonly used when describing processes of non-linear QED: the so-called \textit{classical non-linearity parameter} of the laser field, $\xi=e\mathcal{E}_L/m_e\omega_L$, where $\omega_L$ is the laser frequency, quantifies the coupling of the laser field to the probe charge. In the regime where $\xi\ll1$, QED is fully perturbative, while, as $\xi$ approaches 1, an increasing number of higher-order processes contribute, until for $\xi>1$, the perturbative expansion breaks down and QED becomes non-perturbative in the coupling to the laser field. The second parameter is the \textit{quantum parameter} $\chi_e=E^\star/\mathcal{E}_{\text{crit}}$ which, in case of electron-laser interaction, quantifies the ratio of the effective field strength in the electron rest frame and the critical field. The quantum parameter $\chi_\gamma$ for photon-laser collisions is defined analogously as $\chi_\gamma=(1+\cos\theta)E_\gamma/(m_e \mathcal{E}_{\text{crit}})$.\\

\begin{figure}[h]
     \centering
     \begin{subfigure}[b]{0.49\textwidth}
         \centering
         \includegraphics[width=\textwidth]{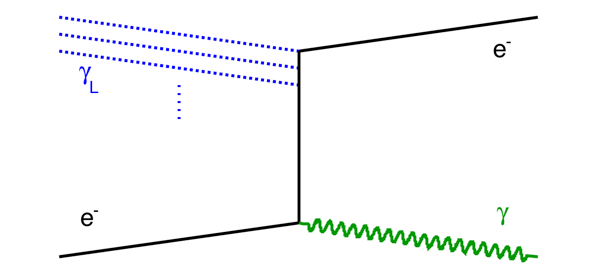}
         \caption{Non-linear Compton scattering.}
         \label{fig:feyncompton}
     \end{subfigure}
     \hfill
     \begin{subfigure}[b]{0.49\textwidth}
         \centering
         \includegraphics[width=\textwidth]{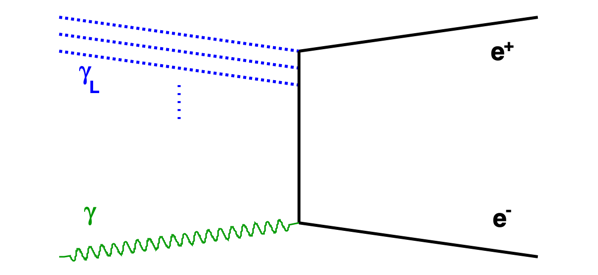}
         \caption{Non-linear Breit-Wheeler pair production.}
         \label{fig:feynbppp}
     \end{subfigure}
     \caption{Feynman diagrams of strong-field QED processes at LUXE~\cite{LUXECDR}.}
\end{figure}

In electron-laser collisions, the process of interest for the LUXE experiment are \textit{non-linear Compton scattering} (see figure \ref{fig:feyncompton}) and subsequent \textit{non-linear Breit-Wheeler pair production} (see figure \ref{fig:feynbppp}). In non-linear Compton scattering, the incident photon absorbs multiple laser photons, emitting a single Compton photon. This Compton photon undergoes non-linear Breit-Wheeler pair creation by absorbing multiple laser photons and creating an electron-positron pair. The combined process $e^- + n\gamma_L \rightarrow e^- + e^+ + e^- $, known also as the \textit{Trident process} can in general  take place as a "one-step" process with an off-shell virtual photon, or as "two step" process with a real photon. In the LUXE parameter regime, the two-step process is dominant, whereas the one-step process is negligible. In photon-laser collisions, the incident high-energy Bremsstrahlung photons undergo non-linear Breit-Wheeler pair production (figure \ref{fig:feynbppp}) directly, in the reaction $\gamma_B + n\gamma_L \rightarrow e^+ + e^- $~\cite{PhysRevLett.26.1072}. \\

The main experimental goal of LUXE is to measure the positron rate of the Trident process and the direct Breit-Wheeler pair production as a function of the laser intensity parameters $\chi$ and $\xi$.

\section{LUXE experimental setup}
\label{sec:luxesetup}
The LUXE experiment will be located at the end of the Eu.XFEL electron accelerator. A dedicated beam-line was designed to extract one single electron bunch out of 2500 bunches per Eu.XFEL train to be guided to the LUXE experimental area at a repetition rate of $10\,\text{Hz}$~\cite{beamlinecdr}. The laser will operate at a $1\,\text{Hz}$ repetition rate, enabling in-situ background and calibration data-taking. This design allows LUXE to collect a large sample of precision data while being transparent to the Eu.XFEL photon science programme. \\

The field strength of the high-intensity laser in the laboratory frame, $\mathcal{E}_L$, with current state-of-the-art lasers, is three orders of magnitude below the Schwinger limit. However, the Lorentz boost of the 16.5 GeV incoming electrons leads to an enhancement of the field $E^\star$ experienced by the electrons in their rest frame:

\begin{equation}
E^\star=\gamma_{\text{e}}\mathcal{E}_L(1+cos{\theta}),
\end{equation}

\noindent
where $\gamma_{\text{e}}\sim10^4$ is the electron Lorentz factor and $\theta=17.2^\circ$ is the foreseen crossing angle between the electron beam and the optical laser pulse in LUXE. As a result, field strengths above the Schwinger limit are accessible at LUXE with current laser technologies.\\

%Figure \ref{fig:luxesetups} shows a schematic of the LUXE electron-laser (\ref{fig:setuptrident}) and the photon-laser  (\ref{fig:setupbppp}) setups with the fully equipped experimental area. 

%There is no strict length limitation, but the authors are strongly encouraged to keep contents to the strict minimum necessary for peers to reproduce the research described in the paper.

%\subsection{Electron-laser collisions}
%You are free to use dividers as you see fit.
%\subsection{Photon-laser collisions}
%Figures should only occupy the stricly necessary space, in any case individually fitting on a single page. Each figure item should be appropriately labeled and accompanied by a descriptive caption. {\bf SciPost does not accept creative or promotional figures or artist's impressions}; on the other hand, technical drawings and scientifically accurate representations are encouraged.

\subsection{The Laser}
\label{sec:laser}

The laser intended for the LUXE experiment is a femtosecond-pulsed Titanium-Sapphire optical laser ($\lambda_L=800\,\text{nm}$) using the chirped-pulse-amplification technique~\cite{Strickland:1985gxr}. A staged approach is foreseen, where for \textit{phase-0} the existing, well-characterized Jenaer Titan:Saphir $40\,\text{TW}$ Laser System (JETI40) will be used, which for \textit{phase-1} of LUXE will be replaced by an upgraded $350\,\text{TW}$ laser system. The laser intensity parameter $\xi$ is varied by changing the laser focal spot size in the interaction point, with the highest $\xi\sim7.9 (23.6)$ in phase-0 (phase-1) achieved with a $3\,\mu m$ spot size. The maximum quantum parameter $\chi_e$ achievable is $\chi_e=1.5 (4.45)$ with the phase-0 (phase-1) setup. Among the most important characteristics of the laser system is its precise shot-to-shot stability. This will be ensured by several redundant diagnostics measurements both on the unperturbed outgoing laser beam itself, as well as by precision determination of particle spectra after the interaction points.
  
\subsection{Particle detectors}
\label{sec:detectors}

Figure \ref{fig:luxesetups} shows a schematic drawing of the particle detectors for the electron-laser and the photon-laser setup. Both setups employ a magnetic dipole spectrometer (B=1.5T) to measure the energy spectrum of electrons and positrons after the interaction point. For every detection area, two redundant detector technologies are foreseen, to enable cross-calibration and reduction of systematic uncertainties. One of the biggest challenge of the LUXE experiment are the vastly different particle rates in different detection areas of the setup. The detector technologies have been specifically adapted for the conditions in which they need to operate in LUXE:

\begin{figure}[h]
     \centering
     \begin{subfigure}[b]{0.49\textwidth}
         \centering
         \includegraphics[width=\textwidth]{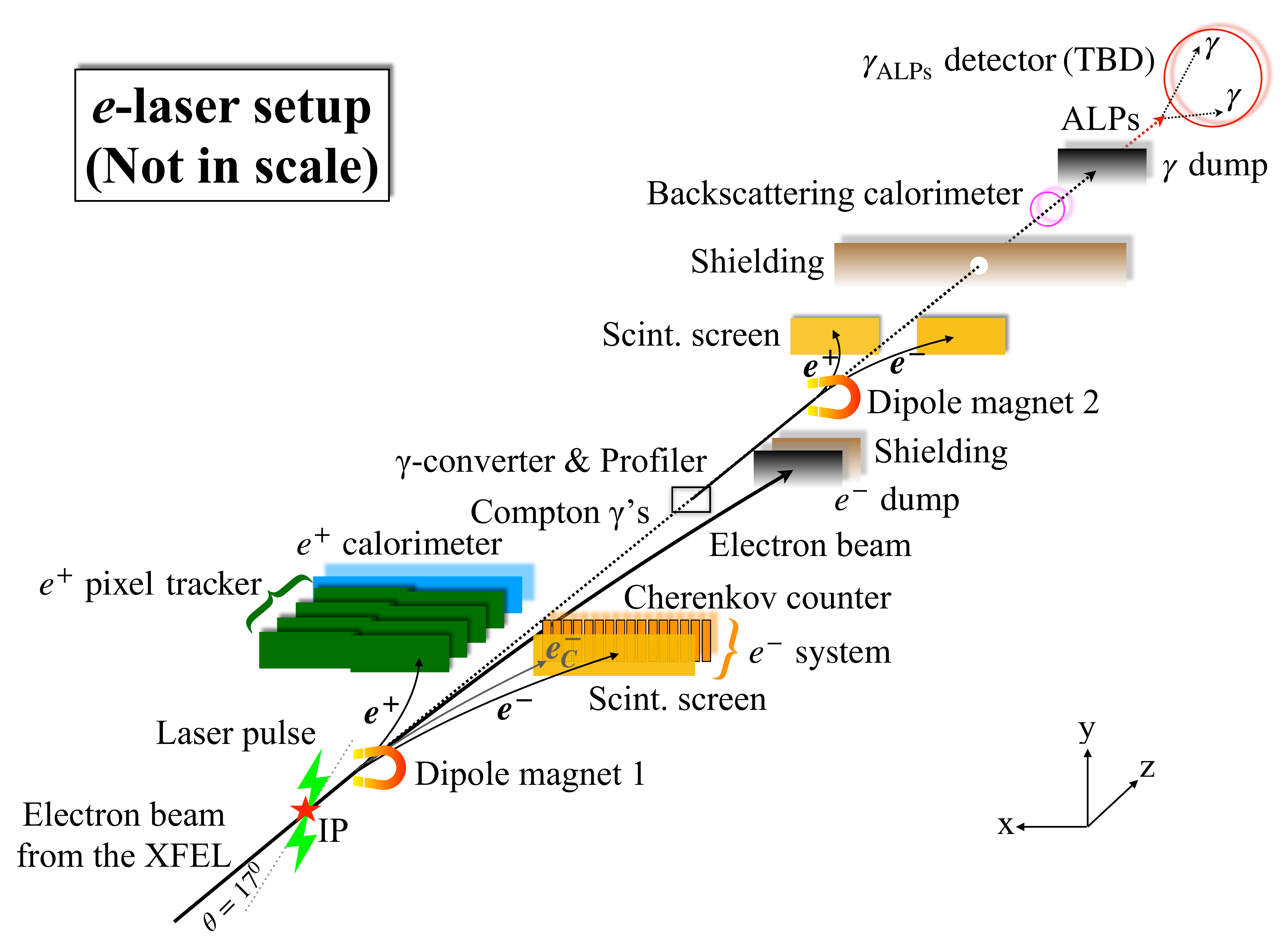}
         \caption{Electron-laser setup.}
         \label{fig:setuptrident}
     \end{subfigure}
     \hfill
     \begin{subfigure}[b]{0.49\textwidth}
         \centering
         \includegraphics[width=\textwidth]{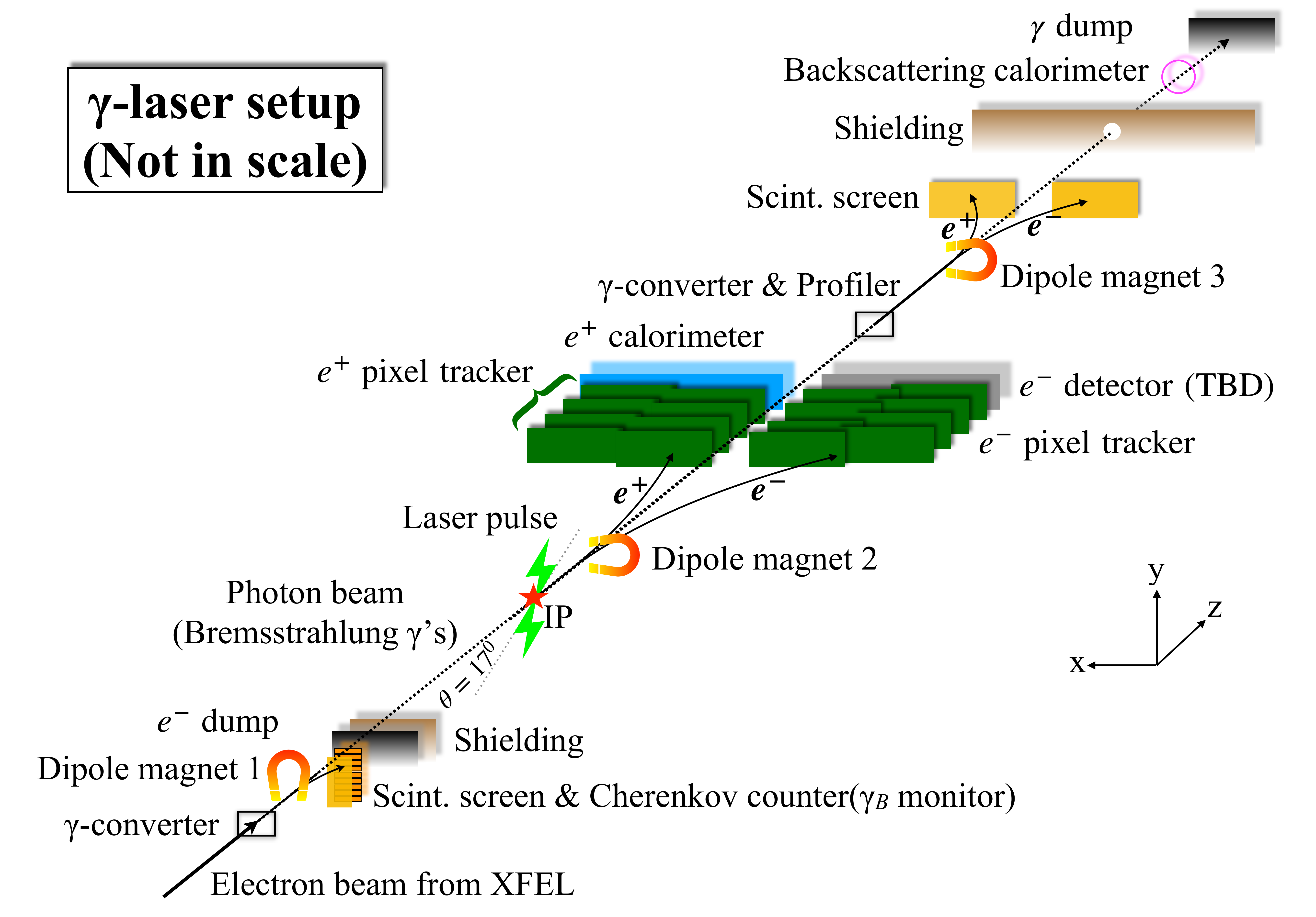}
         \caption{Photon-laser setup.}
         \label{fig:setupbppp}
     \end{subfigure}
     \caption{Sketch of the LUXE experimental setups~\cite{LUXECDR}.}
      \label{fig:luxesetups}	
\end{figure}

\begin{itemize}
\item{\textbf{positron detectors:}} The positron rates for the Breit-Wheeler and Trident process vary between $10^{-3} - 10^4 e^+/\text{shot}$. The foreseen detector system consists of four layers of Silicon pixel tracking detectors using the ALPIDE sensor technology, as well as a high-granularity Tungsten-Silicon electromagnetic calorimeter. The main objectives are a high signal efficiency and good background rejection. 
 
\item{\textbf{electron detectors:}} For the electron-laser setup the rate of Compton-scattered electrons is of the order of $10^9/\text{shot}$, therefore the detectors must be tolerant of such high rates. Foreseen are segmented Argon-filled Cherenkov detectors, measuring the particle rate as a function of deflection by the magnetic field, as well as a Scintillator screen that is read out by an optical camera. The Scintillator provides excellent position resolution, while the Cherenkov detector helps to reject photons and low-energy backgrounds due to the Cherenkov threshold of $\sim 20\,\text{MeV}$. In case of the photon-laser setup, the particle rates between electron and positron side are comparable. In this case the detector system on the electron side will consist also of a Tracker and Calorimeter. 

\item{\textbf{photon detectors:}} In both electron- and photon-laser setups there is a large flux of photons. The photon energy and flux are characterized by several systems, namely by studying the conversion into electron-positron pairs on a target, as well as directly profiling the photon beam angular distribution using Sapphire strips. Finally, the photon flux is determined by measuring backscattering from the photon beam dump in a crystal calorimeter. 

\end{itemize}

\subsection{Parasitical BSM search}
\label{sec:bsmsearch}

\begin{figure}[h]
\centering
\includegraphics[width=0.7\textwidth]{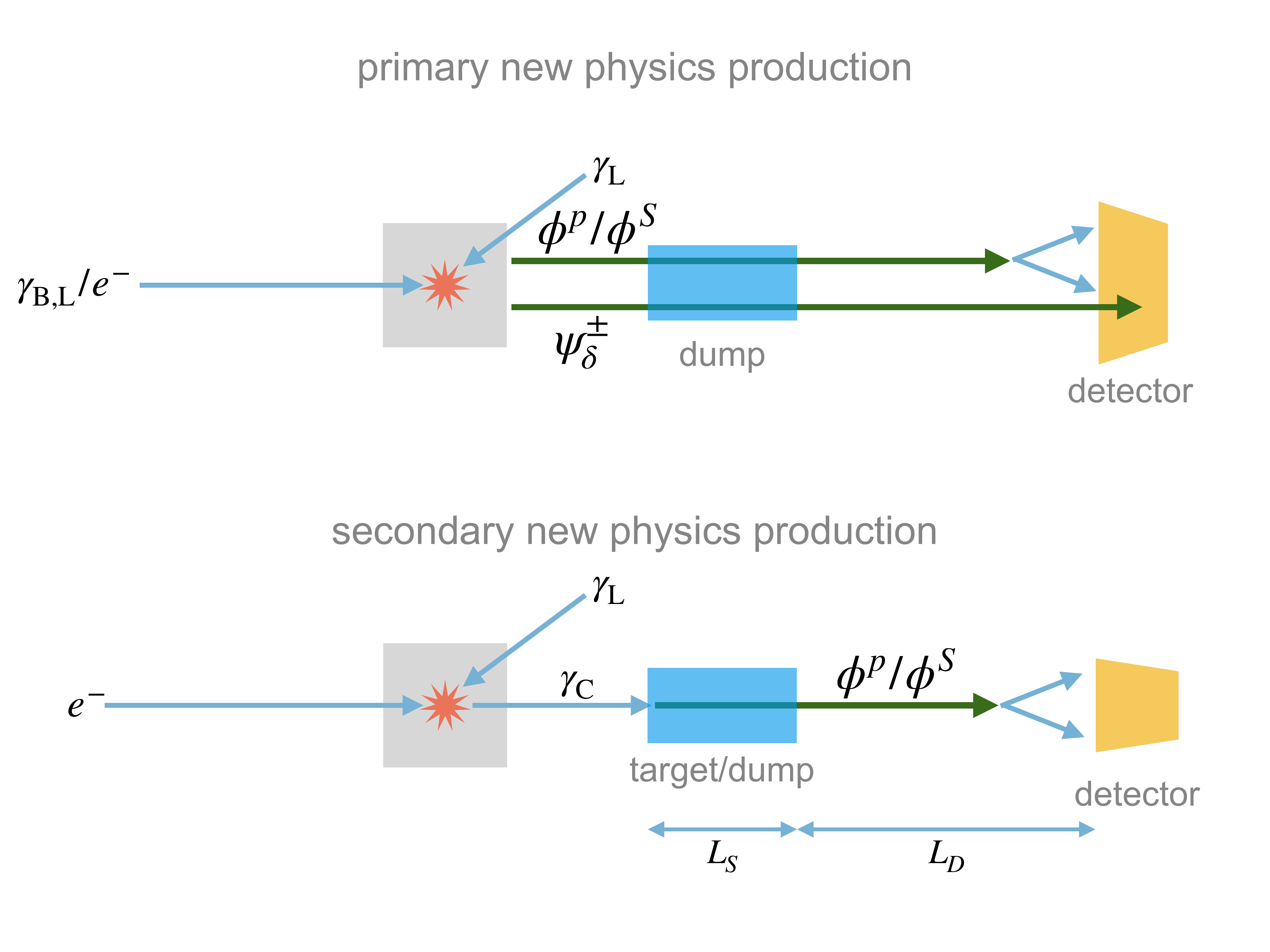}
\caption{Possible production modes of long-lived BSM particles in LUXE~\cite{LUXECDR}.}
\label{ref:bsmscenarios}
\end{figure}

In both the electron-laser and the photon-laser setup, the LUXE experiment produces a high-intensity beam of GeV photons, in the former case via Compton-scattering and in the latter, via Bremsstrahlung. A photon beam dump is foreseen at the end of the LUXE setup. One of the possible scenarios for Beyond-the-Standard Model physics in LUXE is the production of axion-like particles (ALPs) in the photon beam dump material (see figure \ref{ref:bsmscenarios}), which decay into a pair of photons behind the dump and get detected by a photon detector at the end of the LUXE experimental setup. With a $1\times1m^2$ detector area, the expected sensitivity of LUXE to axion-like particles in the lifetime-mass plane is comparable with that of other ongoing and future experiments~\cite{LUXEBSM}. 

\section{Expected results}
\label{sec:expresults}

The main goal of the LUXE experiment is to measure the positron rate from vacuum polarization as a function of the laser intensity parameter $\xi$ and the quantum parameter $\chi$ and to compare it to theoretical predictions from strong-field QED. In strong-field QED, the probability to create an electron-positron pair via Breit-Wheeler pair production, $P$, is expected to have a tunneling-like dependence $P\propto \chi_\gamma\exp\left(-\frac{8}{3\chi_\gamma}\right)$, for $\xi\gg1$. At low $\xi\ll1$, the behavior can be approximated by a power law $P\propto\xi^{2n}$, however the perturbative expansion breaks down for values of $\xi>1$. \\

\begin{figure}[h]
     \centering
     \begin{subfigure}[b]{0.49\textwidth}
         \centering
         \includegraphics[width=\textwidth]{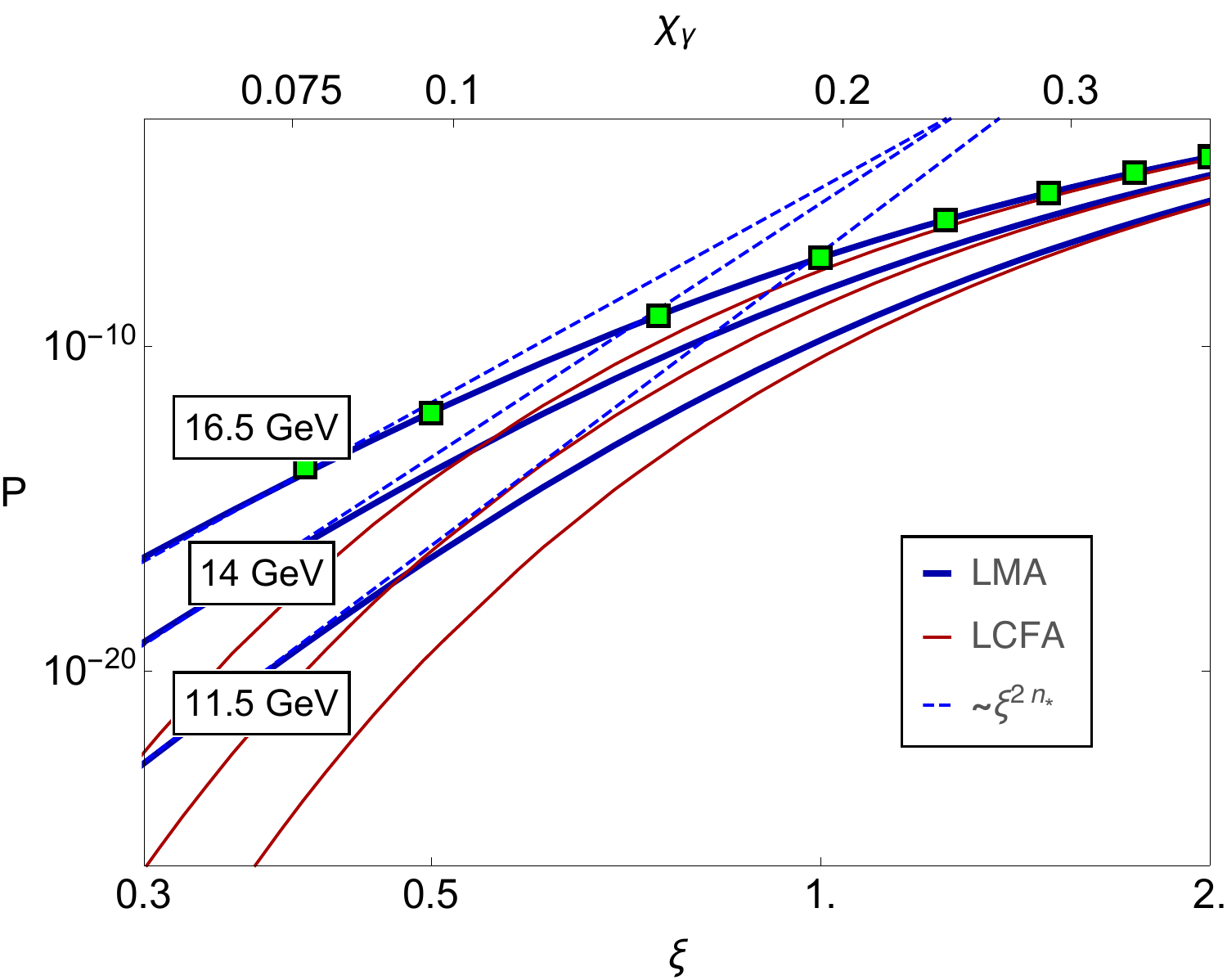}
         \caption{Predicted probability of Breit-Wheeler pair production for the LUXE experiment~\cite{LUXECDR}.}
        \label{fig:bpppprob}
     \end{subfigure}
     \hfill
     \begin{subfigure}[b]{0.49\textwidth}
         \centering
         \includegraphics[width=\textwidth]{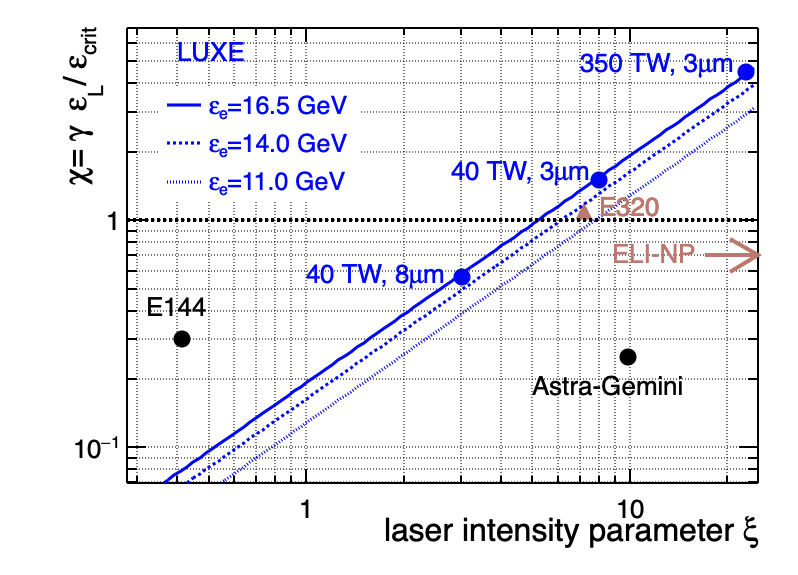}
         \caption{LUXE and other strong-field QED experiments in the strong-field QED parameter space~\cite{LUXECDR}.}
         \label{fig:luxeparamspace}
     \end{subfigure}
     \caption{}
\end{figure}

Figure \ref{fig:bpppprob} shows the expected probability of Breit-Wheeler pair production as a function of the laser intensity parameter $\xi$ and the quantum parameter $\chi$. It is clearly visible how the rate predicted by strong-field QED departs from the power-law expectation from perturbative calculation in the parameter regime that will be reached by LUXE. It is expected that the main source of uncertainty on the positron rate originates from the uncertainty of the laser intensity parameter $\xi$. The goal is to achieve an uncertainty on $\xi$ of $5\%$, which results in an uncertainty of about $40\%$ on the positron production rate. Several complementary measurements of $\xi$ are foreseen in the LUXE setup to ensure that $\xi$ is determined with sufficient precision.

Figure \ref{fig:luxeparamspace} shows the region in parameter space that is accessible by the LUXE experiment, in comparison to a selection of past and future strong-field QED experiments. The $\chi$ and $\xi$ values accessible with LUXE depend on the power of the laser system as well as on the laser spot size. With the foreseen upgrade to a $350\,\text{TW}$ laser power in connection with a small spot size ($3\,\mu\text{m}$), LUXE is expected to reach well into the non-perturbative regime of $\chi>1$ and $\xi>1$. Compared to the previous E144 experiment ~\cite{Bula:1996st,Burke:1997ew}, the LUXE phase-1 setup will attain a factor of 20 higher $\chi$ and a factor of 60 higher $\xi$. This improvement is mainly due to the higher laser intensity of the LUXE laser. Another novelty of the LUXE experiment compared to previous strong-field QED experiments are the foreseen collisions between real, externally produced GeV photons with the laser.

\section{Conclusion}
\label{sec:conclusion}

The LUXE experiment aims to explore a previously uncharted regime of Quantum electrodynamics by studying electron-laser and photon-laser interactions, with the potential of being the first to enter the non-perturbative Schwinger regime of QED. Using one bunch per train of the $16.5\,\text{GeV}$ electron beam of the European XFEL, a state-of the art optical Terawatt laser system and specifically optimized detector technologies, LUXE aims to collect a large dataset of electron- and photon-laser collisions in order to perform precision measurements of strong-field QED. The Compton photon beam of LUXE impinging on a beam dump can be used to search for signs of physics beyond the Standard Model, with the BSM detector running at the same time as the photon- or electron-laser collision data-taking. 

\section*{Acknowledgements}
This work was in part funded by the Deutsche Forschungsgemeinschaft under Germany's Excellence Strategy -- EXC 2121 ``Quantum Universe" -- 390833306.

\bibliography{Biblio.bib}

\begin{thebibliography}{10}
\providecommand{\url}[1]{\texttt{#1}}
\providecommand{\urlprefix}{URL }
\expandafter\ifx\csname urlstyle\endcsname\relax
  \providecommand{\doi}[1]{doi:\discretionary{}{}{}#1}\else
  \providecommand{\doi}{doi:\discretionary{}{}{}\begingroup
  \urlstyle{rm}\Url}\fi
\providecommand{\eprint}[2][]{\url{#2}}

\bibitem{Heisenberg:1935qt}
W.~Heisenberg and H.~Euler,
\newblock \emph{{Consequences of Dirac's theory of positrons}},
\newblock Z. Phys. \textbf{98}, 714 (1936),
\newblock \eprint{arXiv:physics/0605038}.

\bibitem{Schwinger:1951}
J.~Schwinger,
\newblock \emph{{On Gauge Invariance and Vacuum Polarization}},
\newblock Phys. Rev. \textbf{82}, 664 (1951).

\bibitem{Ruffini:2009hg}
R.~Ruffini, G.~Vereshchagin and S.-S. Xue,
\newblock \emph{{Electron-positron pairs in physics and astrophysics: from
  heavy nuclei to black holes}},
\newblock Phys. Rept. \textbf{487}, 1 (2010),
\newblock \doi{10.1016/j.physrep.2009.10.004},
\newblock \eprint{arXiv:0910.0974}.

\bibitem{Nishikov}
A.~I. Nishikov,
\newblock \emph{{Absorption of high-energy photons in the universe}},
\newblock Sov. Phys. JETP \textbf{14}, 393 (1962).

\bibitem{Kouveliotou:1998ze}
C.~Kouveliotou \emph{et~al.},
\newblock \emph{{An X-ray pulsar with a superstrong magnetic field in the soft
  gamma-ray repeater SGR 1806-20.}},
\newblock Nature \textbf{393}, 235 (1998).

\bibitem{Turolla:2015mwa}
R.~Turolla, S.~Zane and A.~Watts,
\newblock \emph{{Magnetars: the physics behind observations. A review}},
\newblock Rept. Prog. Phys. \textbf{78}, 116901 (2015),
\newblock \doi{10.1088/0034-4885/78/11/116901},
\newblock \eprint{arXiv:1507.02924}.

\bibitem{Kaspi:2017}
V.~M. Kaspi and A.~M. Beloborodov,
\newblock \emph{Magnetars},
\newblock Annual Review of Astronomy and Astrophysics \textbf{55}, 261 (2017),
\newblock \doi{10.1146/annurev-astro-081915-023329}.

\bibitem{LUXECDR}
{LUXE collaboration},
\newblock \emph{{Conceptual Design Report for the LUXE Experiment}} (2021),
  \eprint{arXiv:2102.02032}.

\bibitem{PhysRevLett.26.1072}
H.~R. Reiss,
\newblock \emph{Production of electron pairs from a zero-mass state},
\newblock Phys. Rev. Lett. \textbf{26}, 1072 (1971),
\newblock \doi{10.1103/PhysRevLett.26.1072}.

\bibitem{beamlinecdr}
F.~Burkart and W.~Decking,
\newblock \emph{{Extraction and XTD20 Transfer Line: Conceptual Design
  Report}}.

\bibitem{Strickland:1985gxr}
D.~Strickland and G.~Mourou,
\newblock \emph{{Compression of amplified chirped optical pulses}},
\newblock Opt. Commun. \textbf{55}, 447 (1985),
\newblock [Erratum: Opt. Commun. 56 (1985) 219].

\bibitem{LUXEBSM}
Z.~Bai, O.~Davidi, A.~Hartin, T.~Ma, G.~Perez, Y.~Soreq, A.~Santra and
  N.~Tal~Hod,
\newblock \emph{{in progress}} .

\bibitem{Bula:1996st}
C.~Bula \emph{et~al.},
\newblock \emph{{Observation of nonlinear effects in Compton scattering}},
\newblock Phys. Rev. Lett. \textbf{76}, 3116 (1996).

\bibitem{Burke:1997ew}
D.~L. Burke \emph{et~al.},
\newblock \emph{{Positron production in multi - photon light by light
  scattering}},
\newblock Phys. Rev. Lett. \textbf{79}, 1626 (1997).

\end{thebibliography}

\nolinenumbers

\end{document}